\documentclass[conference]{IEEEtran}
\IEEEoverridecommandlockouts
\usepackage{cite}
\usepackage{amsmath,amssymb,amsfonts}
\usepackage{algorithmic}
\usepackage{graphicx}
\usepackage{textcomp}
\usepackage{xcolor}
\usepackage{booktabs}
\usepackage{makecell}
\usepackage{subfigure}
\def\BibTeX{{\rm B\kern-.05em{\sc i\kern-.025em b}\kern-.08em
    T\kern-.1667em\lower.7ex\hbox{E}\kern-.125emX}}
\begin{document}

\title{Exploring Modality Disruption in Multimodal Fake News Detection}

\author{
	\IEEEauthorblockN{
		Moyang Liu$^{1}$, 
		Kaiying Yan$^{2}$, 
		Yukun Liu$^{3*}$\thanks{*Corresponding Author.}, 
		Ruibo Fu$^{4}$ 
		Zhengqi Wen$^{5}$
        Xuefei Liu$^{4}$
        Chenxing Li$^{6}$} 
	\IEEEauthorblockA{$^{1}$Beihang University, Beijing, China}
	\IEEEauthorblockA{$^{2}$School of Mathematics, Sun Yat-sen University, Guangzhou, China}
	\IEEEauthorblockA{$^{3}$School of Artificial Intelligence, University of Chinese Academy of Sciences, Beijing, China} 
	\IEEEauthorblockA{$^{4}$Institute of Automation, Chinese Academy of Sciences, Beijing, China}
        \IEEEauthorblockA{$^{5}$Beijing National Research Center for Information Science and Technology, Tsinghua University, Beijing, China}
        \IEEEauthorblockA{$^{6}$Tencent AI Lab}
    \IEEEauthorblockA{\{moyang\_liu\}@buaa.edu.cn, \{yanky\}@mail2.sysu.edu.cn, \{yukunliu927\}@gmail.com}
} 

\maketitle

\begin{abstract}
The rapid growth of social media has led to the widespread dissemination of fake news across multiple content forms, including text, images, audio, and video. Compared to unimodal fake news detection, multimodal fake news detection benefits from the increased availability of information across multiple modalities. However, in the context of social media, certain modalities in multimodal fake news detection tasks may contain disruptive or over-expressive information. These elements often include exaggerated or embellished content. We define this phenomenon as modality disruption and explore its impact on detection models through experiments. To address the issue of modality disruption in a targeted manner, we propose a multimodal fake news detection framework, FND-MoE. Additionally, we design a two-pass feature selection mechanism to further mitigate the impact of modality disruption. Extensive experiments on the FakeSV and FVC-2018 datasets demonstrate that FND-MoE significantly outperforms state-of-the-art methods, with accuracy improvements of 3.45\% and 3.71\% on the respective datasets compared to baseline models.
\end{abstract}


\section{Introduction}
\label{sec:intro}
In recent years, with the rapid growth of social media platforms, the spread of fake news has emerged as a significant global societal issue. On social media, users can rapidly disseminate information through likes, shares, and comments, regardless of its veracity\cite{shu2017fake}. Such fake news can not only mislead public perception and influence social opinion\cite{zhou2020survey}, but it can also have severe impacts on various domains, including politics, economics, and public health\cite{zhang2020overview}.

As a result, accurately detecting fake news becomes crucial in helping individuals identify and differentiate between authentic and false content\cite{shu2019beyond}. Nowadays, fake news on social media appears in various forms, such as text, images, audio, and video, making unimodal detection methods insufficient\cite{hu2024bad}. Text can be manipulated to bypass keyword filters\cite{jain2018fake}, images and videos can be altered, and audio can be synthesized or edited\cite{zhang2025cross}. Thus, relying on unimodal detection often fails to deliver ideal results\cite{liu2024exploring}.

To address these challenges, multimodal fake news detection\cite{bu2024fakingrecipe} has emerged. Multimodal detection methods integrate information from multiple sources, such as text, images, audio, and video, utilizing cross-modal feature fusion and correlation analysis to achieve a more comprehensive identification of fake news\cite{nan2024let}. Song et al.\cite{song2021multimodal} proposed a multimodal fake news detection framework based on Cross-modal Attention Residual and Multi-channel convolutional neural Network (CARMN). However, these feature fusion methods often encounter the issue of losing information at the shallow layers\cite{chen2024metasumperceiver}. In response to this issue, Jing et al.\cite{jing2023multimodal} proposed a network that can capture the representational information of each modality at different levels. To cover all possible modalities encountered on social media, Qi et al.\cite{qi2023fakesv} constructed China's largest fake news short video dataset, FakeSV, which includes various contents such as titles, videos, keyframes, audios, metadata, and user comments. 

However, in the context of multimodal fake news detection on social media, we have discovered a novel phenomenon called modality disruption. Specifically, in this context, the content posted by users often includes certain modalities that contain disruptive or excessively expressive information. For example, in the text modality, users frequently employ exaggerated, rhetorical, or even sarcastic titles\cite{alonso2021sentiment}. Similarly, in the video modality, exaggerated special effects or disjointed editing may be present\cite{chang2023fake}. These elements are typically designed to attract attention through hyperbolic embellishments. Although such expressions do not alter the inherent truthfulness label of the news, they may influence the decision-making process of fake news detection models to some extent. Nevertheless, no prior research has explored the impact of such data characteristics in this specific context. Existing approaches typically incorporate all modality information into the fake news detection task without considering whether it is reasonable to include certain modalities in the model's decision-making process\cite{alghamdi2024comprehensive}.

To explore the phenomenon of modality disruption in multimodal fake news detection within the context of social media, we analyzed the data characteristics specific to social media platforms and conducted a series of related experiments. And it can be confirmed that modality disruption impairs the performance of detection models. To address this issue in a targeted manner,  we have developed a framework for multimodal fake news detection called Fake News Detection Mixture of Experts (FND-MoE). The framework adopts a two-pass feature selection mechanism, effectively mitigating modality disruption by excluding disruptive or excessively expressive embellishments in modality information. Furthermore, after ensuring the resolution of modality disruption, the framework introduces distinct feature extractors for each modality to fully capture the complete information within each, further improving detection accuracy and robustness.

Our contributions can be summarized in three aspects:

\begin{itemize}
    \item We explored the phenomenon of modality disruption in multimodal fake news detection within the context of social media, and through a series of experiments, we confirmed its negative impact on the performance of detection models.
    \item We proposed a framework specifically designed to address the modality disruption issue and demonstrated its effectiveness through experiments.
    \item We designed a two-pass feature selection mechanism to mitigate the impact of modality disruption, effectively improving the accuracy of multimodal fake news detection.
\end{itemize}

\section{Modality Disruption}

Modality disruption refers to the presence of disruptive, overly expressive, or ironic information within certain modalities of a published news piece. This characteristic is particularly prevalent in social media contexts, where users often employ exaggerated embellishments to attract attention. For instance, in the text modality, users tend to craft sensational, rhetorically charged headlines designed to capture interest; in the video modality, videos may include exaggerated special effects or disjointed editing. These embellishments do not alter the inherent truthfulness of the news itself, but they can disrupt the model's ability to accurately assess the content's authenticity.

Fake news often seeks to mislead by distorting or misinterpreting information within certain modalities\cite{liu2024skepticism}. In addition to assessing the factual credibility of each modality, existing multimodal fake news detection methods typically rely on assessing the consistency and matching of information across different modalities to determine the authenticity of the news\cite{hettiachchi2023designing}. Although the embellishments mentioned earlier do not change the inherent truthfulness of the news, they interfere with the model's ability to assess the relationships between modalities accurately. This modality disruption phenomenon ultimately affects the model's capacity to make correct predictions regarding the authenticity of the news.

\begin{figure}[htb]
	\centering
	\subfigure[]{
		\begin{minipage}[t]{0.5\linewidth}
			\centering
			\includegraphics[width=1.75in]{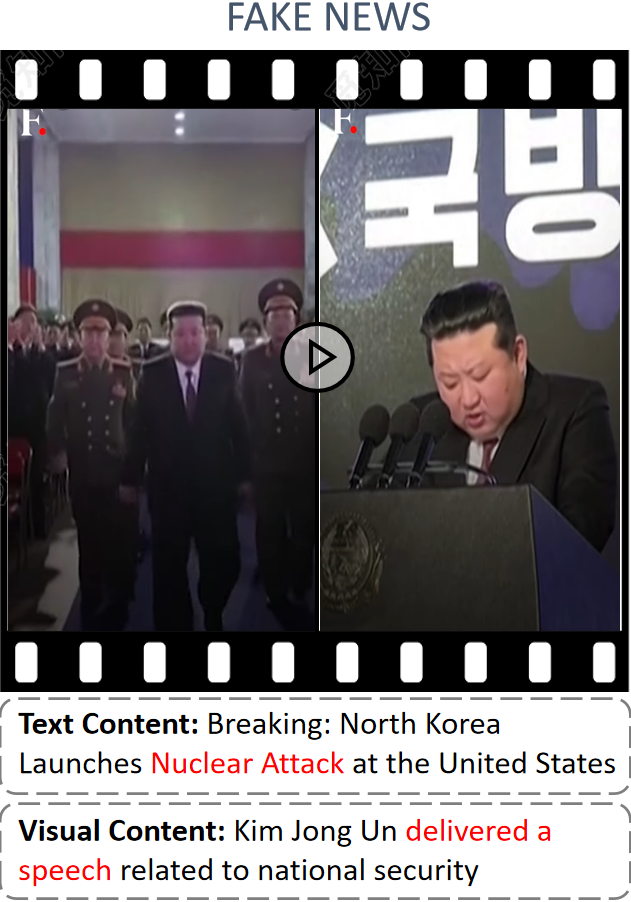}\\
		\end{minipage}%
	}%
	\subfigure[]{
		\begin{minipage}[t]{0.5\linewidth}
			\centering
			\includegraphics[width=1.75in]{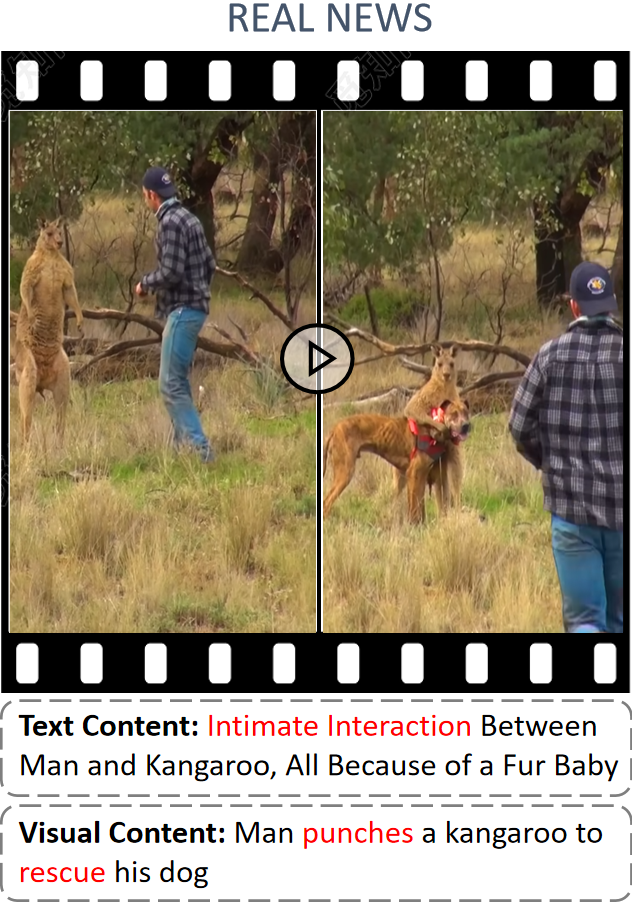}\\
		\end{minipage}%
	}%
	\centering
	\caption{Potential scenarios that may arise on social media. (a) illustrates that fake news often aims to mislead by manipulating and distorting factual information within a specific modality to create a deceptive narrative. (b) demonstrates that some modalities contain disruptive or overly expressive embellishments designed to attract attention, which does not alter the inherent truthfulness of the news.}
	\vspace{-0.2cm}
	\label{fig2}
\end{figure}

For example, Fig. \ref{fig2} (a) highlights that fake news creators often seek to mislead the public by distorting factual information within certain modalities, such as manipulating headlines or maliciously editing videos. Consequently, existing fake news detection methods primarily focus on evaluating the consistency of information across different modalities to assess the authenticity of the news. Fig. \ref{fig2} (b) further illustrates that while some news content may be factually accurate, publishers often employ disruptive, exaggerated, or ironic embellishments within certain modalities to capture attention. In such cases, the model may struggle to assess the news's authenticity accurately, leading to potential misjudgments.

A series of experiments were conducted to conclusively demonstrate the existence of the modality disruption issue, which is particularly severe in the context of multimodal fake news detection. The discussion will be elaborated in detail in the experiment section.

\begin{figure*}[h]
    \centering
    \setlength{\abovecaptionskip}{-0cm}
    \includegraphics[width=17cm]{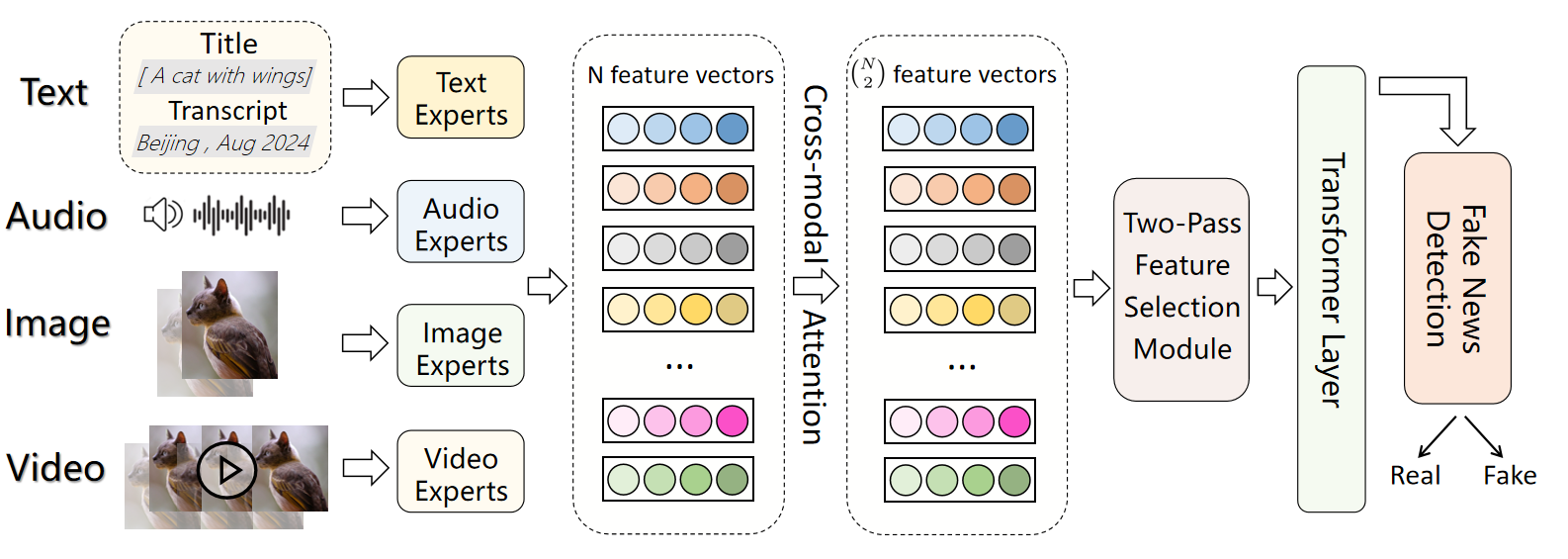}
    \caption{Architecture of the multimodal fake news detection framework FND-MoE }
    \label{f1}
\end{figure*}

\section{Methods}

\subsection{Framework}

To address the modality disruption issue in multimodal fake news detection, we propose a novel framework, FND-MoE, as shown in Fig. \ref{f1}. 
In this framework, to tackle the widespread occurrence of various types of misinformation encountered in real-world scenarios, we leverage multiple modalities within short video content—including text, audio, image, and video, to comprehensively extract relevant information. For each modality, distinct expert models are employed to extract features and integrate the extracted modal information for fake news detection. 
Next, we apply cross-attention\cite{vaswani2017attention} between the features extracted by the expert models for each modality pairwise, aiming to capture the complementary information across different modalities. Assuming there are \( N\) modality experts, performing pairwise cross-attention on the \( N\) extracted features would yield $\binom{N}{2}$ modality-enhanced vectors. Subsequently, these modality-enhanced features undergo a two-pass feature selection mechanism. This two-pass feature selection mechanism comprises an attention-based top-k gating filter\cite{shazeer2017outrageously} and a Gumbel-Sigmoid-based\cite{jang2016categorical} dynamic selection mechanism. Finally, the selected features \(F_{final}\) are fed into a transformer layer, and the output of this layer is classified into true or false categories by a classifier:
\begin{equation}
    \hat{y} = \text{Classifier}(\text{Transformer}(F_{\text{final}}))
\end{equation}

\subsection{Multimodal Feature Extractors}

In order to extract valuable information from each modality, it is crucial to select appropriate expert models based on the current modality.

In the text modality, BERT\cite{devlin2018bert} is a mature and highly versatile text encoder. Therefore, we utilize BERT as the expert model to extract textual features. For the audio modality, we employed two expert models: the CNN-based VGG\cite{simonyan2014very} and the transformer-based Wav2Vec2.0\cite{baevski2020wav2vec} for feature extraction. This choice is motivated by VGG's strength in capturing background audio features, while Wav2Vec2.0 excels at extracting semantic information\cite{fan2020exploring}. By leveraging these complementary models, we aim to achieve a comprehensive understanding of the audio content across different levels. To detect the spatiotemporal and multi-granularity information in the video modality, we used a pre-trained C3D\cite{tran2015learning} model to extract motion features. For the image modality, we extracted a specific number of frames from each video and fed them into a pre-trained VGG19\cite{simonyan2014very} model to learn static visual features.

\begin{figure}[htb]
    \centering
    \setlength{\abovecaptionskip}{-0.3cm}
    \includegraphics[width=0.5\textwidth]{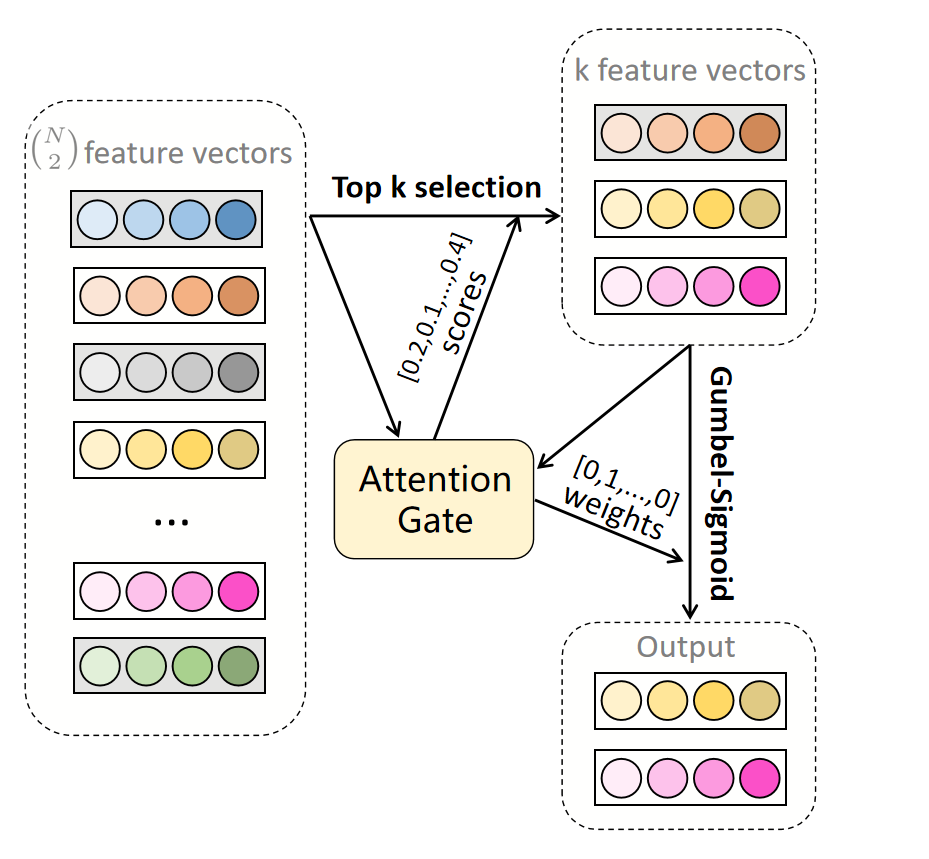}
    \caption{Architecture of two-pass feature selection mechanism. The input set of $\binom{N}{2}$ feature vectors is processed through an attention gate, from which the top k vectors are selected based on their scores. These k feature vectors are then passed through another attention gate, where the Gumbel-Sigmoid function is applied to generate the final output.}
    \label{f2}
\end{figure}

\begin{table*}[ht]
\centering
\renewcommand\arraystretch{1.1}
\setlength{\belowcaptionskip}{0.2cm}
\caption{The experimental results of the multimodal fake news detection framework using unimodal features and different combinations of modalities (text, audio, image, video ). Combinations 1-7 represent a series of experiments conducted within the SV-FEND framework, each utilizing different modalities for feature extraction and fusion. Combination 1 and Combination 2 used different audio encoders, VGG and Wav2Vec2.0, respectively. Combination 7 used VGG and Wav2Vec2.0 encoders simultaneously. }
\label{table1}
\begin{tabular}{ccccccccccccc}
\Xhline{1.2pt}
           & \multicolumn{4}{c}{Modality} & \multicolumn{4}{c}{FakeSV}                               & \multicolumn{4}{c}{FVC-2018}                                      \\ \hline
Method     & Text & Audio & Image & Video & Acc            & F1             & Pre            & Rec   & Acc            & F1             & Pre            & Rec            \\ \hline
VGGish+SVM &      & \checkmark     &       &       & 61.25          & 61.31          & 61.24          & 61.33 & 58.44          & 58.61          & 58.48          & 58.63          \\
VGG19+Att  &      &       & \checkmark     &       & 68.53          & 68.51          & 68.53          & 68.50 & 65.79          & 65.81          & 65.49          & 66.08          \\
C3D+Att    &      &       &       & \checkmark     & 70.26          & 70.24          & 70.25          & 70.25 & 71.81          & 71.72          & 71.89          & 71.85          \\
Bert+Att   & \checkmark    &       &       &       & 74.31          & 74.35          & 74.30          & 74.39 & 76.37          & 76.35          & 76.39          & 76.33          \\ \hline
Combination 1         & \checkmark    & \checkmark     &       &       & 81.55          & 81.40          & 81.34          & 81.73 & 78.61          & 78.20          & 78.21          & 78.54          \\
Combination 2         & \checkmark    & \checkmark     &       &       & 82.29          & 81.03          & 83.44          & 81.21 & 79.77          & 79.50          & 79.36          & 79.43          \\
Combination 3         & \checkmark    & \checkmark     & \checkmark     &       & 78.60          & 78.46          & 78.40          & 78.78 & 84.39          & 84.01          & 84.32          & 84.13          \\
Combination 4         & \checkmark    & \checkmark     &       & \checkmark     & 78.60          & 78.96          & 78.80          & 79.57 & 83.82          & 84.16          & 83.27          & 84.50          \\
Combination 5         & \checkmark    &       & \checkmark     & \checkmark     & 79.84          & 79.12          & 78.53          & 78.83 & 86.12          & 84.77          & 85.58          & 84.25          \\
Combination 6         &      & \checkmark     & \checkmark     & \checkmark     & 78.60          & 78.03          & 78.21          & 77.58 & 81.50          & 81.57          & 81.60          & 81.63          \\
Combination 7         & \checkmark    & \checkmark     & \checkmark     & \checkmark     & 78.12          & 78.83          & 78.60          & 79.33 & 83.82          & 83.98          & 83.72          & 84.89          \\ \Xhline{1.2pt}

\end{tabular}
\vspace{-0cm}
\end{table*}

\subsection{Two-pass Feature Selection Mechanism}

Compared to unimodal fake news detection, multimodal fake news detection benefits from an increased amount of available information due to the addition of multiple modalities. However, in social media contexts, some disruptive modality features may influence the model's ability to make accurate predictions.

Therefore, we designed a two-pass feature selection mechanism, incorporating an attention-based top-k gating filter and a Gumbel-Sigmoid-based dynamic selection mechanism. 

First, we concatenate the \( N\) feature vectors obtained through the cross-attention mechanism: \(F_{\text{concat}} = [F_{1 \to 2}, F_{1 \to 3}, \dots, F_{N-1 \to N}]\).
Subsequently, these feature vectors are fed into a self-attention-based feedforward layer, where each vector is assigned a score:

\begin{equation}
    \alpha_i = \dfrac{\exp(\text{score}(F_i))}{\sum_{j=1}^{N} \exp(\text{score}(F_j)} 
\end{equation}

The top k vectors, based on their scores, are then selected for retention 
\begin{equation}
    F_{\text{top-k}} = \text{TopK}([F_{\text{concat}}])
\end{equation}
ensuring that only the most informative features are preserved for further processing. 

At this point, the Gumbel-Sigmoid technique is applied to the selected top-k features. For each feature, the Gumbel noise is added to the logit score, and the Sigmoid function is used to generate a probability of retaining or discarding the feature. The outcome of this process is the final set of features:
\begin{equation}
    F_{\text{final}} = \{f_i \mid z_i \geq 0.5\}
\end{equation}

This allows the model to exclude the disruptive features that may contain highly embellished content, thereby eliminating modality disruption and improving the overall detection accuracy.

\subsection{Gumbel-Sigmoid Technique}

Gumbel-Sigmoid is a stochastic technique designed for differentiable sampling from discrete distributions. It merges the Gumbel-Max trick\cite{jang2016categorical}, commonly used for categorical sampling, with the Sigmoid function to enable continuous and differentiable approximations of binary decisions. This makes it particularly well-suited for tasks such as feature selection, where we need to decide whether to retain or discard a feature in a way that still allows backpropagation for model optimization during training.

In this mechanism, a Gumbel noise is sampled from a Gumbel distribution, known for modeling the maximum of a set of variables. By applying the noise to a logit (log-odds), followed by a temperature-scaled Sigmoid function, the mechanism approximates a discrete decision in a differentiable manner. Mathematically, the Gumbel noise is generated as:
\begin{equation}
    g_i = -\log(-\log(U_i)), \quad U_i \sim \text{Uniform}(0, 1)
\end{equation}
Then, the final decision for feature selection is computed as:
\begin{equation}
z_i = \sigma\left(\frac{1}{\tau} \left( \log(\alpha_i) + g_i \right)\right)
\end{equation}
where $\sigma$ is the Sigmoid function, and $\tau$ is the temperature parameter. By adjusting $\tau$, we can fine-tune the balance between exploration and exploitation in feature selection.

By using the Gumbel-Sigmoid technique, we ensure that the model can properly discard or retain features during training while still allowing for normal backpropagation to optimize the model.

\begin{table*}[h]
\centering
\renewcommand\arraystretch{1.1}
\setlength{\belowcaptionskip}{0.2cm}
\caption{The experimental results regarding the performance comparison of FND-MoE with other models.}
\label{table1}
\begin{tabular}{ccccccccc}
\Xhline{1.2pt}
           & \multicolumn{4}{c}{FakeSV}                               & \multicolumn{4}{c}{FVC-2018}                                      \\ \hline
Method      & Acc            & F1             & Pre            & Rec   & Acc            & F1             & Pre            & Rec            \\ \hline
TikTec\cite{shang2021multimodal}        & 75.07          & 75.04          & 75.18          & 75.07 & 77.02          & 73.95          & 74.24          & 73.67          \\
FANVN\cite{choi2021using}         & 75.04          & 75.02          & 75.11          & 75.04 & 85.81          & 85.32          & 85.20          & 85.44          \\
SV-FEND\cite{qi2023fakesv}       & 79.31          & 79.24          & 79.62          & 79.31 & 84.71          & 85.37          & 84.25          & 86.53          \\ \hline
FND-MoE      & \textbf{82.84} & \textbf{82.22} & \textbf{83.76} & \textbf{81.60} & \textbf{88.44} & \textbf{89.02} & \textbf{88.56} & \textbf{89.62} \\ \Xhline{1.2pt}

\end{tabular}
\vspace{-0cm}
\end{table*}

\section{Experiments}

\subsection{Datasets and Experiments Details}

We used two multimodal datasets in the context of social media, FakeSV and FVC-2018. The details of the datasets are described as follows:
\begin{itemize}
    \item FakeSV dataset, constructed by Qi et al.\cite{qi2023fakesv}, comprises a large collection of Chinese news short videos. This dataset includes multiple modalities such as text, video, audio, and social context, which can cover various data in social media scenarios.
    \item The FVC-2018 dataset\cite{hu2021fvc} contains real and fake videos on topics from YouTube like politics, sports, and entertainment. It includes multiple modalities such as titles, videos, comments, and URLs, making it valuable for fake news detection.
\end{itemize}

In the experiments, the dataset was divided into training, validation, and test sets in a 70:15:15 ratio following a chronological order. The model utilized the cross-entropy loss function and AdamW optimizer, with a batch size of 64. The final results were obtained by evaluating this best model on the test set.

Qi et al.\cite{qi2023fakesv} proposed a multimodal fake news detection framework, SV-FEND, which integrates text, audio, image, and video modalities based on an attention mechanism. We conducted a series of experiments on feature extraction and fusion based on SVFEND, investigating the model's performance when different modal information is included separately.

\subsection{Experiment Results of Modality Disruption}

The experiments indicate that incorporating additional modal information is not always beneficial. 

First, as shown in Table \ref{sec:intro}, relying solely on unimodal fake news detection can already achieve satisfactory results.

Similarly, it can be observed from Table \ref{sec:intro} that multimodal fake news detection generally outperforms single-modal approaches by a significant margin. This indicates that the complementarity between different modalities can, to a certain extent, enhance the model's performance, as integrating multiple modalities enables a more comprehensive capture of the characteristics of multimodal information.

The results also showed that, on the FakeSV dataset, the fusion of text features and audio features extracted via the wav2vec encoder achieved the best results. Meanwhile, on the FVC-2018 dataset, the optimal model performance was observed when text, video, and image features were fused. It is worth noting that both of these optimal modality feature fusion methods only incorporate a subset of the four available modalities. Other experiments presented in Table \ref{sec:intro} further show that adding more modal information to these fusion strategies results in a decline in performance. This phenomenon highlights the modal disruption issue and the lack of complementarity among the modalities.

\subsection{Performance Results}

As shown in Table \ref{table1}, the experiments demonstrate that our FND-MoE framework outperforms state-of-the-art methods. To address the issue of modality disruption, this framework eliminates disruptive modality features that may contain highly embellished content during the fusion process. After resolving the modality disruption, the framework introduces additional expert models for each modality, enabling a comprehensive capture of modal information at various levels. Therefore, the robustness and accuracy of the detection model are significantly improved. Specifically, compared to the corresponding state-of-the-art methods, SV-FEND, our FND-MoE model achieved accuracy improvements of 3.45$\%$ on the FakeSV dataset and 3.71$\%$ on the FVC-2018 dataset.

\subsection{Ablation Study}

We designed a series of ablation experiments to validate the effectiveness of the model. Specifically, we compared the model's performance under several different gating mechanisms.

First, we examined the differences in model performance and modality selectivity when using sigmoid, softmax, and Gumbel-sigmoid for the final output weighting. It was observed that the model's performance with softmax was inferior to that with Gumbel-Sigmoid. This is because the softmax mechanism exhibited stronger continuity in modality selection, resulting in a smoother weight distribution across modalities and a lack of clear selectivity. In contrast, when employing the Gumbel-Sigmoid mechanism, the model demonstrated more discrete modality selection, effectively distinguishing the contributions of different modalities and thereby enhancing overall model performance. This suggests that the Gumbel-Sigmoid mechanism is more advantageous for processing multimodal information, as it strengthens the model's sensitivity to critical modalities while excluding all of the disruptive modality information, ultimately improving decision-making performance.

\begin{table}[h]
\centering

\renewcommand\arraystretch{1}
\caption{The performance comparison of the model when employing sigmoid, softmax, and Gumbel-Sigmoid mechanisms}
\label{table2}
\setlength{\belowcaptionskip}{0.2cm}
\renewcommand\arraystretch{1.1}
\begin{tabular}{ccccc}
\Xhline{1.2pt}
Mechanism      & Acc            & F1             & Pre            & Rec            \\ \hline
sigmoid        & 81.37          & 81.30          & 81.56          & 81.23          \\
softmax        & 81.89          & 81.67          & 81.95          & 81.43          \\
Gumbel-Sigmoid & \textbf{82.48} & \textbf{82.04} & \textbf{82.30} & \textbf{81.60} \\ \Xhline{1.2pt}
\end{tabular}

\end{table}

\begin{figure}[htb]
    \centering
    \includegraphics[width=0.5\textwidth]{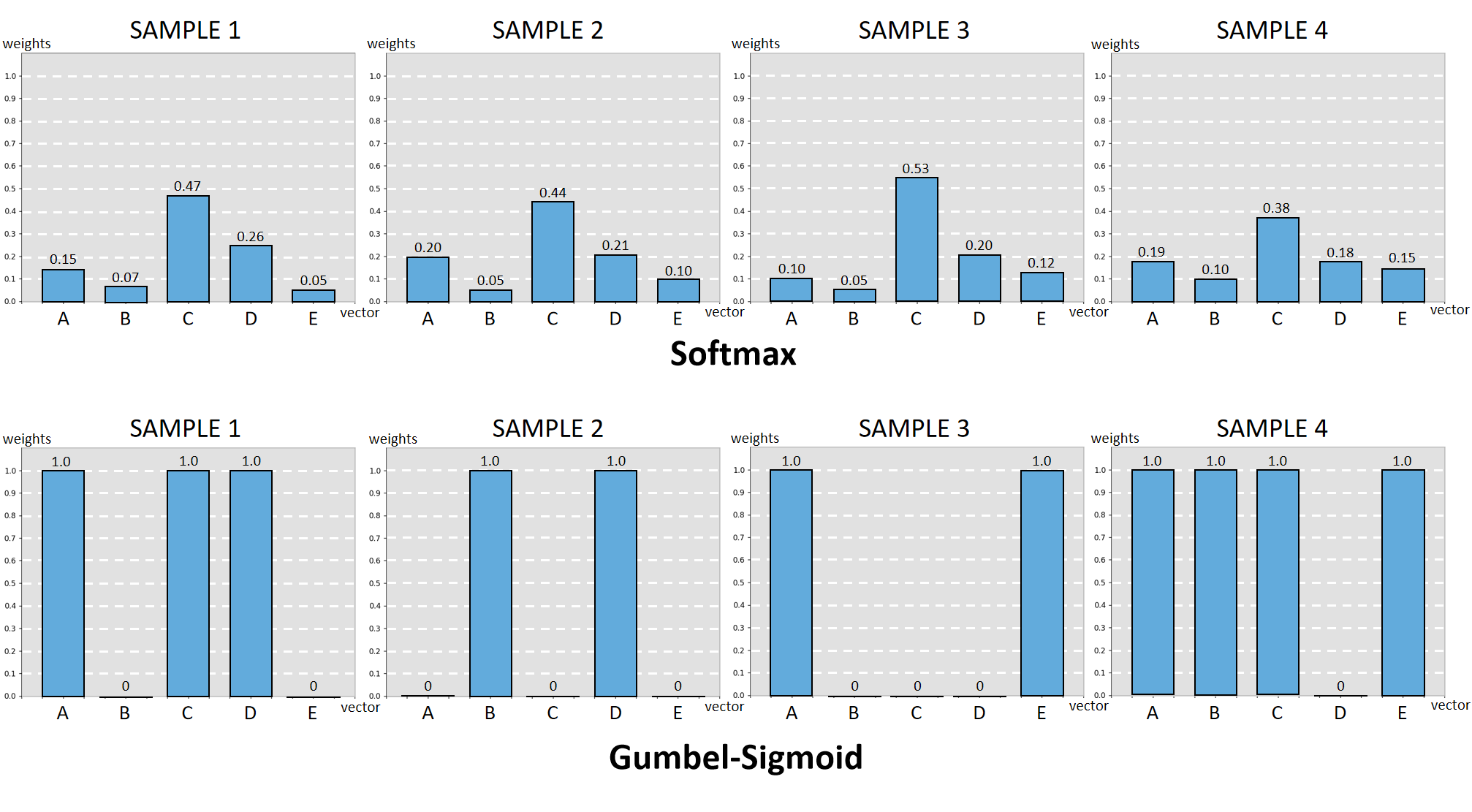}
    \caption{The weight values output by the model when using two mechanisms.}
    \label{f3}
\end{figure}

To validate this modality selection capability, we specifically examined the weight values output by the model when using these two mechanisms. As shown in Fig. \ref{f3}, with the softmax mechanism, the model's output weights tend to follow a fixed pattern. In contrast, when using the Gumbel-Sigmoid mechanism, the selected modalities exhibit no clear regularity in the model's output. This is precisely due to the more discrete nature of the Gumbel-Sigmoid mechanism, which allows the model to dynamically adjust its reliance on different modalities under varying inputs, thus preventing any single modality from disproportionately influencing the model's decisions.

\begin{table}[h]
\centering
\caption{The performance comparison of the model when employing top-k, Gumbel-Sigmoid, and a combination of both mechanisms.}
\label{table3}
\setlength{\belowcaptionskip}{0.2cm}
\renewcommand\arraystretch{1.1}
\begin{tabular}{ccccc}
\Xhline{1.2pt}
Mechanism              & Acc            & F1             & Pre            & Rec            \\ \hline
Top k                  & 80.44          & 80.39          & 80.25          & 80.36          \\
Gumbel-Sigmoid         & 82.48          & 82.04          & 82.30          & 81.60          \\
Top k + Gumbel-Sigmoid & \textbf{82.84} & \textbf{82.22} & \textbf{83.76} & \textbf{81.60} \\ \Xhline{1.2pt}
\end{tabular}
\end{table}

Next, we compared three selection mechanisms: top-k, Gumbel-Sigmoid, and a combination of top-k and Gumbel-Sigmoid, based on attention gating. It is evident that the combination of top-k + Gumbel-Sigmoid yields the best performance. This is because the combination of top-k and Gumbel-Sigmoid effectively leverages the strengths of both mechanisms. The top-k mechanism ensures that only the most relevant modalities are considered by focusing on the top-ranked ones, thereby reducing the disruptive modality information. Meanwhile, the Gumbel-Sigmoid mechanism introduces a level of stochasticity and discrete selection, allowing the model to explore different modality combinations and prevent over-reliance on specific modalities. By combining these two mechanisms, the model filters out the disruptive modality features and dynamically adjusts its modality selection, leading to improved robustness and overall performance.

\section{Conclusion}

In this paper, we explored the modality disruption phenomenon in social media, particularly in multimodal fake news detection. By exploring the characteristics of data in the social media context and conducting a series of experiments, we have confirmed that modality disruption can impact the model's ability to make accurate predictions. To address this issue, we proposed a novel framework, FND-MoE, for multimodal fake news detection that leverages expert models for each modality and a two-pass feature fusion mechanism. Our approach dynamically excludes the disruptive modality features through a combination of top-k gating and Gumbel-Sigmoid mechanisms. Extensive experiments conducted on the FakeSV and FVC-2018 datasets demonstrated that FND-MoE outperforms state-of-the-art methods, achieving significant improvements in accuracy. Additionally, ablation studies were conducted to ensure the effectiveness of our proposed model. Future work can focus on further exploring the characteristics of data on social media and developing targeted solutions to address these challenges.

\bibliographystyle{IEEEbib}
\bibliography{main}

\vspace{12pt}

\end{document}